\documentclass[preprint,eqsecnum,aps,showpacs]{revtex4}
\usepackage{bm} 
\usepackage[dvips]{graphicx}
\usepackage{epsfig}
\begin{document}
\title{Quantum and Thermal Corrections to a Classically
Chaotic Dissipative System}
\author{Manuel Rodr\'{\i}guez-Achach\cite{manuel},
Gabriel P\'erez}
\affiliation{Departamento de F\'{\i}sica Aplicada,
Cinvestav del IPN, Unidad M\'erida,
A. P. 73 ``Cordemex'', 97310 M\'erida, Yucat\'an, M\'exico.}
\author{Hilda A. Cerdeira}
\affiliation{International Centre for Theoretical Physics
Trieste, Italy.}

\begin{abstract}
The effects of quantum and thermal corrections on the dynamics of a damped
nonlinearly kicked harmonic oscillator are studied. This is done via
the Quantum Langevin Equation formalism working on a truncated moment
expansion of the density matrix of the system. We find that the 
type of bifurcations present in the system change upon quantization and
that chaotic behavior appears for values of the nonlinear parameter
that are far below the chaotic threshold for the classical
model. Upon increase of temperature or Planck's constant,
bifurcation points and chaotic thresholds are shifted towards
lower values of the nonlinear parameter.
There is also an anomalous
reverse behavior for low values of the cutoff frequency. 
\end{abstract}
\pacs{05.45.+b, 03.65.Sq, 05.40.+j, 42.50.Lc}

\maketitle

 

\section{Introduction}

With the impressive results obtained in the theory of classical chaos,
the interest in the manifestations of chaos in quantum 
systems has grown
over the past years. Although there is not a clear difference between 
chaotic and regular dynamics for 
quantized Hamiltonian dynamics,
some insight has been gained with the quantization of conservative models
with chaotic classical counterparts. Among the peculiar phenomenology that
has been found
we have the appearance of {\it scars}, which are wavefunctions with 
enhanced amplitude
along the paths of classical unstable periodic orbits
\cite{scar1,scar2,scar3,scar4}.
Further connection between classically unstable orbits and and quantum 
behavior has been given 
by Gutzwiller's trace formula for the density of states\cite{gutz2,gutz}, 
which has been applied with 
some success to many hyperbolic systems \cite{period-orb}.
A connection has also been found between the nature of the 
classical dynamics, be it regular or chaotic, 
and the statistics of the energy spectrum, 
which obeys the general distributions found
in Random Matrix Theory \cite{rand-mat1,rand-mat2}.

However, for systems that include dissipation,
not much work has been done.
This is an important subject of study
because, besides its theoretical interest, it is of central importance
in several interesting applications (SQUIDS, tunneling phenomena, etc.)
The main problem with such systems is the presence
of the dissipative term, which makes the quantization process
not feasible by standard methods.
Approaches to this problem have been proposed by many
authors over the last three decades. Kostin proposes the use of a 
nonlinear Schr\"odinger
equation for the process of quantization \cite{kostin},
but it has some problems with the superposition principle
and yields results of an unphysical nature. Dekker \cite{dekker}
uses a canonical quantization method based on complex variables, with
some degree of success, but the physical basis of his theory is
not very clear. Maximal
Entropy Methods \cite{kowalski} have also been proposed,
using a coupling between quantum and classical degrees of freedom.
This approach ignores the effects of thermally driven fluctuations.

In general, the only fully successful  and physically sound
quantization scheme for dissipative systems is the so-called 
system--plus--reservoir approach, in which one considers
the system coupled to a large reservoir or heath bath.
Since the entire system plus reservoir may be considered as closed,
one can apply to it the standard rules of quantization, paying of course the
penalty of dealing with an infinite number of degrees of freedom.
Dissipation comes about by the net transfer of energy from the
system to the reservoir, and thermal fluctuations appear as a result
of the random energy exchange from/to the heat bath.
These approaches ensure the non-violation of commutator rules, and include
the appropriate fluctuation--dissipation relations.

Some common approaches using the system-plus-reservoir picture
are the influence-functional methods
introduced by Feynman and Vernon \cite{feynman2},
that have been used by Caldeira and Legget to study quantum 
Brownian motion \cite{caldeira1} and dissipative tunneling 
effects \cite{caldeira2};
generalized master equations \cite{master},
used to describe damping phenomena in quantum optics and spin
dynamics \cite{optics,haken75};
and Quantum Langevin Equations \cite{ford88}, used by Ford 
{\em et al.\/}
to study dissipative quantum tunneling at zero temperature \cite{ford}.

In the area of nonlinear (and classically chaotic) dissipative models,
there are some results that point to the conclusion that
quantum effects act as sources of noise. As a result, 
it is expected that 
the threshold for chaos be lowered {\em in a continuous fashion}
upon the introduction 
of quantum corrections. This was reported
by Graham and Tel \cite{graham_tel}, who worked in the master equation
formalism \cite{haken75,gardiner91} to quantize Henon's map, and by Cerdeira
{\em et al.\/} \cite{cerdeira1}, who built a quantum mapping 
of Wigner functions for a kicked
oscillator.
More recently, Cohen \cite{cohen}
studied the quantum dissipative kicked rotator using influence functionals,
while Dittrich and Graham used the master equation formalism
for the same purpose  \cite{dittrich1}. 
Liu and Schieve \cite{schieve} use
the Quantum Langevin Equation approach to show that a chaotic attractor
is observed in a quantum system whose classical counterpart is non-chaotic.
In contradiction to 
these works, there were some results
showing an increase of order (a raising of the chaoticity thresholds) 
for a kicked dissipative oscillator \cite{balasundaram1}. 

In this work we study the behavior of a periodically kicked harmonic
oscillator, with dissipation, and in the presence of quantum 
and thermal fluctuations. Our interest is mostly to find how the 
bifurcation points and chaotic thresholds in parameter space
change when increasingly large thermal and quantum effects are included.
To do this, we generate a mapping of expectation values, 
using a quantum Langevin
approach, that has an immediate analog with its classical
limit. This Expectation Values Mapping (EVM) constitutes in principle a 
complete description of the quantum dynamics, based in a discrete set
of variables, and has the advantage of exhibiting a clear hierarchy
in its terms, introducing from the beginning mean values and a sequence
of moments of the fluctuations (the completeness of the description
given by the EVM will be shown in appendix~A).

In section II we introduce the classical model, a harmonic oscillator
with a periodic ``kick" from a nonlinear potential. In section III
we will use the Langevin equation formalism
to obtain a quantal version of the classical map of 
the preceding section. The
resulting (truncated) EVM will include both thermal and quantum corrections;
the change in the behavior of the dynamics as function of $k_BT$ and $\hbar$
is examined. In section IV and V we show and discuss
the results obtained and give some concluding remarks.

\section{Classical Kicked Harmonic Oscillator with Damping}

Starting with the Hamiltonian for a kicked oscillator
\begin{equation}
\label{H_clas}
H=\frac{p^2}{2m}+\frac{1}{2}m\Omega^2x^2+V(x)\sum_n\delta(t-n\tau),
\end{equation}
where $\tau$ is the period of the kick,
we introduce the dissipative force $2\gamma\dot x$ to obtain
the equations of motion in the form
\begin{eqnarray}
\label{ec_a_integrar}
\dot p &=&-m\Omega^2x-2\gamma\frac{p}{m}-V'(x)\sum_n\delta(t-n\tau),\\
\dot x &=&\frac{p}{m}.
\end{eqnarray}
From now on we will set the units of mass and time so that $m=1$
and $\Omega=1$.
For times $t\neq n\tau$, the kick has no effect so we get the equation
\begin{equation}
\ddot x+2\gamma\dot x+x=0,
\end{equation}
whose well known underdamped ($\gamma < 1$) solution is given by
\begin{eqnarray}
\label{cont1}
x(t)&=&e^{-\gamma t}(A\cos\omega_0 t+B\sin\omega_0 t),\\
\label{cont2}
p(t)&=&e^{-\gamma t}\left[\omega_0(-A\sin\omega_0 t+B\cos\omega_0 t)
-\gamma(A\cos\omega_0 t+B\sin\omega_0 t)\right],
\end{eqnarray}
where $\omega_0 = \sqrt{1-\gamma^2}$, and $A$ and $B$ are given in
terms of initial conditions.
Evaluating these constants, and defining an evolution matrix $T$ by
\begin{eqnarray}
T_{QQ}& \equiv & \exp(-\gamma \tau) (\cos(\omega_0 \tau) 
+\gamma \sin(\omega_0 \tau) / \omega_0),\\
T_{QP}& \equiv &\exp(-\gamma \tau) \sin(\omega_0 \tau) /\omega_0,\\
T_{PQ}& \equiv &-\exp(-\gamma \tau) (\omega_0+\gamma^2/\omega_0)
\sin(\omega_0 \tau),\\
T_{PP}& \equiv &\exp(-\gamma \tau) (\cos(\omega_0 \tau) -\gamma 
\sin(\omega_0 \tau)/\omega_0),
\end{eqnarray}
we can write a discrete linear map corresponding to 
Eqs.~(\ref{cont1}, \ref{cont2})
\begin{eqnarray}
x_{n+1}^- &=& T_{QQ} \, x^+_n + T_{QP} \, p^+_n,\\
p_{n+1}^- &=& T_{PQ} \, x^+_n + T_{PP} \, p^+_n,
\end{eqnarray}
where we have set as usual 
$x_n^- \equiv x(n \tau - 0^+)$, 
$x_n^+ \equiv x(n \tau + 0^+)$, with
similar definitions for $p$.
We need now to calculate how the kick affects the dynamics,
in order to complete the map. This can
be done by integrating Eq.~(\ref{ec_a_integrar}) from
just before to just after
$t=(n+1)\tau$. Over this region
$x$ is continuous and thus we obtain
\begin{eqnarray}
\label{en1}
p^+_{n+1}-p_{n+1}^- &=& -V'(x^-_{n+1}),\\
\label{en2}
x^+_{n+1} &=& x_{n+1}^-.
\end{eqnarray}
Successive application of the mappings (\ref{cont1}, \ref{cont2}) and 
(\ref{en1}, \ref{en2})
will give us a map that yields the values for $x$ and $p$ just after
each kick (dropping from now on the superscripts):
\begin{eqnarray}
x_{n+1} &=& T_{QQ} \, x_n + T_{QP} \, p_n,\\
p_{n+1} &=& T_{PQ} \, x_n + T_{PP} \, p_n - V'(x_{n+1}).
\end{eqnarray}

We have chosen the force acting on the kicks as
\begin{equation}
\label{potencial}
V'(x) = -V_0e^{-(x/a)^2},
\end{equation}
which gives a potential bounded from
below. From now on we will set the units of length so that $a=1$.
This force is maximum at $x = 0$, so that
the closer the harmonic oscillator gets to its rest position, the
stronger the repulsive effect of the kick becomes.
We are interested in a region of
parameter space such that $\gamma \ll \omega$, 
and where clear chaotic behavior can be found.
After some search, we have chosen $\gamma = 0.03$
and $\tau = 10.0$. For bifurcation 
diagrams given in terms of $V_0$, we find a well 
defined transition to chaos around $V_0 \approx 5.69287$.
It should be noticed that there is some coexistence of attractors in 
 this sector of  parameter space; 
these attractors evolve from order to chaos following
Feigenbaum scenario (see Fig.~(1)). For the fully chaotic region,
the attractors are H\'enon-like (see Fig.~(2)).

The later map will be quantized in the next section in order to see
the temperature and quantum effects on the dynamics.

\section{Quantization of the Kicked Harmonic Oscillator with Damping}

We use the Langevin equation formalism for the process of
quantization.   
This equation is obtained from the quantization in the 
Heisenberg picture of a closed system given by a relevant degree of freedom 
$\mbox{\boldmath$Q$}$, 
in contact with a large (in the limit, infinite)
bath of harmonic oscillators $\mbox{\boldmath$q$}_j$. 
It is given by
\begin{equation}
\label{lang_eq}
\mbox{\boldmath$\ddot Q$}+2\int_{t_e}^t\Gamma(t-t')
\mbox{\boldmath$\dot Q$}dt'+W'(\mbox{\boldmath$Q$})=
\mbox{\boldmath$F$}(t),
\end{equation}
where $\mbox{\boldmath$Q$}(t)$ and $\mbox{\boldmath$F$}(t)$ are
operators, $W(\mbox{\boldmath$Q$})$ is the potential and
$\mbox{\boldmath$F$}(t)$ is a random force that, together with the
memory function $\Gamma(t)$, depends on the characteristics of the bath.
These oscillators are assumed to be in thermal equilibrium at 
some initial time $t_e$.
The memory function $\Gamma(t)$, 
and the expectation value and symmetric correlations of 
$\mbox{\boldmath$F$}(t)$ are given by \cite{ford88}
\begin{eqnarray}
\langle\mbox{\boldmath$F$}(t)\rangle&=&0,\\  \label{autocorr}
\langle\mbox{\boldmath$F$}(t)\mbox{\boldmath$F$}(t')+\mbox{
\boldmath$F$}(t')\mbox{\boldmath$F$}(t)\rangle&=&
\sum_j \hbar m_j \omega_j^3 
\coth(\frac{\hbar\omega_j}{2k_BT}) \cos[\omega_j(t-t')],\\
\Gamma(t) &=& \frac{1}{2}\sum_j m_j \omega_j^2 \cos(\omega_j t) \Theta(t).
\end{eqnarray}
where $m_j$ and $\omega_j$ are the mass and frequency of the $j$th 
oscillator. These expressions imply a quantum fluctuation-dissipation 
theorem.
It is assumed that no correlation
exists between $\mbox{\boldmath$F$}$ and $\mbox{\boldmath$Q$}$
or $\mbox{\boldmath$P$}$. This implies ignoring some switch-on
processes, which arise due to our assumption that the bath {\em per se}
is in thermal equilibrium at time $t_e$.

From now on, we will use a bath of equal masses and a
white ({\em i.~e.}, $\omega$--independent), and continuous
frequency distribution, which is the case for constant
friction or Markovian approximation. In this case, 
\begin{eqnarray}
\Gamma(t)&=&2 \gamma \delta(t), \\
\langle\mbox{\boldmath$F$}(t)\mbox{\boldmath$F$}(t')+\mbox{
\boldmath$F$}(t')\mbox{\boldmath$F$}(t)\rangle&=&
\frac{2\gamma}{\pi}\int_o^\infty d\omega\,\hbar\omega\coth
\left(\frac{\hbar\omega}{2k_BT}\right)\cos[\omega(t-t')],
\end{eqnarray}
and Eq.~(\ref{lang_eq}) reduces to the simpler form
\begin{equation}
\label{q_eq}
\mbox{\boldmath$\ddot Q$}+2\gamma
\mbox{\boldmath$\dot Q$}+W'(\mbox{\boldmath$Q$})=
\mbox{\boldmath$F$}(t).
\end{equation}
The drawback of using the Markovian approximation is that it will
generate some divergences since we now have
contributions from arbitrarily high frequencies in the bath.
We will use the same harmonic oscillator-plus-kick potential
indicated in the Hamiltonian (\ref{H_clas}) and Eqn.~(\ref{potencial}).
Between kicks, the (underdamped) solution of Eq.~(\ref{q_eq}) is given by
\begin{eqnarray}
\mbox{\boldmath$Q$}(t)&=&e^{-\gamma t}(\mbox{\boldmath$A$}\cos\omega_0 t
+\mbox{\boldmath$B$}\sin\omega_0 t)+\mbox{\boldmath$f$}(t,\omega_0,\gamma),\\
\mbox{\boldmath$P$}(t)&=&\mbox{\boldmath$\dot Q$}(t),
\end{eqnarray} 
where the solution for the non--homogeneous part has been obtained from a
Riccatti equation and we have defined
$\omega_0$ in an analogous way as for the classical case. 
The operators
$\mbox{\boldmath$A$}$ and
$\mbox{\boldmath$B$}$ are given in terms of initial conditions,
evaluated at some initial time $t_0$ (not to be confused with $t_e$), and
the inhomogeneous term is 
\begin{equation}
\mbox{\boldmath$f$}(t,\omega_0,\gamma)=\frac{1}{\omega_0}\int_{t_0}^t dt'
e^{-\gamma(t-t')}\sin\omega_0(t-t')\mbox{\boldmath$F$}(t').
\end{equation}
For the moment $\mbox{\boldmath$P$}$ we get
\begin{eqnarray}
\mbox{\boldmath$P$}(t)&=&-\gamma e^{-\gamma t}\left[\mbox{\boldmath$A$}
\cos(\omega_0t)+\mbox{\boldmath$B$}\sin(\omega_0t)\right]\\ 
&&+ \omega_0e^{-\gamma t}\left[-\mbox{\boldmath$A$}\sin(\omega_0t)+
\mbox{\boldmath$B$}\cos(\omega_0t)\right]+
\mbox{\boldmath$g$}(t,\omega_0,\gamma),
\end{eqnarray}
where
\begin{equation}
\mbox{\boldmath$g$}(t,\omega_0,\gamma)=\frac{1}{\omega_0}\int_{t_0}^tdt'
e^{-\gamma(t-t')}\left[-\gamma\sin\omega_0(t-t')+\omega_0\cos\omega_0(t-t')
\right]\mbox{\boldmath$F$}(t').
\end{equation}
Now we will take the integration limits to be
$t_0\to n\tau^+$ and $t\to(n+1)\tau^-$,
so we can evaluate the initial conditions and get
\begin{eqnarray}
\mbox{\boldmath$Q$}^-_{n+1} & = &
     T_{QQ}\, \mbox{\boldmath$Q$}_n^++T_{QP}\, \mbox{\boldmath$P$}_n^+
     +\mbox{\boldmath$f$}(\tau,\omega_0,\gamma),\\
\mbox{\boldmath$P$}^-_{n+1} & = &
     T_{PQ}\, \mbox{\boldmath$Q$}_n^++T_{PP}\, \mbox{\boldmath$P$}_n^+
     +\mbox{\boldmath$g$}(\tau,\omega_0,\gamma),
\end{eqnarray}
with the same matrix elements $T_{ij}$ defined before.
Defining the fluctuations as
\begin{equation}
\delta\mbox{\boldmath$X$} \equiv
\mbox{\boldmath$X$}-\langle\mbox{\boldmath$X$}\rangle,
\end{equation}
we can now write down the evolution equations for expectation values
and fluctuations:
\begin{eqnarray}
\label{Qmap}
\left(\begin{array}{c}
Q_{n+1}^- \\
P_{n+1}^-\end{array}\right)&=&
\left(\begin{array}{cc}
T_{QQ} & T_{QP} \\
T_{PQ} & T_{PP}\end{array}\right)
\left(\begin{array}{c}
Q_n^+\\
P_n^+\end{array}\right),\\
\label{mat_fluc}
\left(\begin{array}{c}
\delta\mbox{\boldmath$Q$}_{n+1}^- \\
\delta\mbox{\boldmath$P$}_{n+1}^-\end{array}\right)&=&
\left(\begin{array}{cc}
T_{QQ} & T_{QP} \\
T_{PQ} & T_{PP}\end{array}\right)
\left(\begin{array}{c}
\delta\mbox{\boldmath$Q$}_n^+\\
\delta\mbox{\boldmath$P$}_n^+\end{array}\right)+
\left(\begin{array}{c}
\mbox{\boldmath$f$}(\tau,\omega_0,\gamma)\\
\mbox{\boldmath$g$}(\tau,\omega_0,\gamma)\end{array}\right),
\end{eqnarray}
since both $\mbox{\boldmath$f$}$ and $\mbox{\boldmath$g$}$
have zero expectation value.
Hereafter we will use non-bold letters to denote expectation values. In
order to find what happens over the kick, we note that $\mbox{\boldmath$Q$}$
is continuous and $\mbox{\boldmath$P$}$ has an increment, so
\begin{eqnarray}
\mbox{\boldmath$Q$}^+_{n+1}&=&
     \mbox{\boldmath$Q$}^-_{n+1} = \mbox{\boldmath$Q$}_{n+1},\\
\label{taylor}
\mbox{\boldmath$P$}^+_{n+1}&=&\mbox{\boldmath$P$}^-_{n+1}-V'(
     \mbox{\boldmath$Q$}_{n+1}).
\end{eqnarray}
Since the potential term has an operator argument we will
perform a second order Taylor expansion. This is the first truncation we 
need to carry out in order to close the system of equations. From this we get
\begin{equation}
\mbox{\boldmath$P$}^+_{n+1}=\mbox{\boldmath$P$}^-_{n+1}
     -V'(Q_{n+1})-V''(Q_{n+1})\delta\mbox{\boldmath$Q$}_{n+1}
     -\frac{V'''(Q_{n+1})}{2}(\delta\mbox{\boldmath$Q$}_{n+1})^2.
\end{equation}
The potential and its derivatives can now be evaluated
numerically since we have no longer
an operator as argument. However in order to have an EVM 
suitable for numerical work, as in Eq.\ (\ref{Qmap}),
we need to take expectation values of all operators.
In order to do this we have to calculate, among others,
the map for the fluctuation
$(\delta Q_{n+1})^2$ so that we can evaluate the above equation for
$P_{n+1}^+$. We proceed
by evaluating $(\delta Q_{n+1})^2$ and made another
truncation to second order in $\delta$, so that we get a map
which is also second order.
In an analogous fashion, we obtain the mappings for the fluctuations
$(\delta P_{n+1}^+)^2$ and
$(\delta P_{n+1}^+\, \delta Q_{n+1}+\delta Q_{n+1}\, \delta P_{n+1}^+)$
and after some rearrangement of terms we have the 5
expectation value maps for $Q$ and $P$
and for the fluctuations (some details are given in Appendix~B)
\begin{eqnarray}
\label{map_begin}
Q_{n+1}&=&T_{QQ}\, Q_n+T_{QP}\, P_n^+,\\
P_{n+1}^+&=&T_{PQ}\, Q_n+T_{PP}\, P_n^+-V'(Q_{n+1})-\frac{V'''(Q_{n+1})}{2}
           (\delta Q_{n+1})^2\\
(\delta Q_{n+1})^2&=&T_{QQ}^2\, (\delta Q_n)^2+T_{QP}^2\, (\delta P_n^+)^2\\
\nonumber
&&+T_{QQ}\, T_{QP}\, (\delta Q_n\, \delta P_n^++\delta P_n^+\, \delta Q_n)
+f\cdot f,\\
(\delta P_{n+1}^+)^2&=&R_{PQ}^2\, (\delta Q_n)^2+R_{PP}^2\, (\delta P_n^+)^2\\
\nonumber
&&+R_{PQ}\, R_{PP}\, 
(\delta Q_n\, \delta P_n^++\delta P_n^+\, \delta Q_n)+h_{n+1}\cdot h_{n+1},\\
\label{map_end}
\delta P_{n+1}^+\, \delta Q_{n+1}+\delta Q_{n+1}\, \delta P_{n+1}^+&=&
2R_{PQ}\, T_{QQ}\,
(\delta Q_n)^2+2T_{QP}\, R_{PP}\, (\delta P_n^+)^2\\
\nonumber
&&+(R_{PQ}\, T_{QP}+R_{PP}\, T_{QQ})
(\delta Q_n\, \delta P_n^++\delta P_n^+\, \delta Q_n)\\
\nonumber
&&+(h_{n+1}\cdot f+f\cdot h_{n+1}).
\end{eqnarray}
Here the following definitions have been made:
\begin{eqnarray}
h_n\cdot f+f\cdot h_n&=&\langle\mbox{\boldmath$h$}_n\cdot\mbox{\boldmath$f$}
+\mbox{\boldmath$f$}\cdot\mbox{\boldmath$h$}_n\rangle,\\
\mbox{\boldmath$h$}_n&=&\mbox{\boldmath$g$}-V''(Q_n)\mbox{\boldmath$f$},\\
R_{PQ}&=&T_{PQ}-V''(Q_{n+1})T_{QQ},\\
R_{PP}&=&T_{PP}-V''(Q_{n+1})T_{QP},
\end{eqnarray}
and cubic and higher order combinations of $h_n$, $f$, $\delta P$ and
$\delta Q$ have been neglected. In principle, a complete description
of the quantum map could be given if one could use all moments 
$\langle P_mQ_m\rangle$. As an extra approximation,
a cutoff frequency has to be
introduced in order to calculate expressions which depend on 
Eq.~(\ref{autocorr}), where the
divergences introduced by the use of the Markovian approximation
show up.
We will now use this set of equations to investigate the effects of
quantum and thermal contributions to the dynamics of the system.
The parameters and potential used are the same of the
classical map.

\section{Results}

In this section we present the results obtained using the maps
developed in the previous section. We analyze the shifts that
occur in the threshold of chaos for the system as function
of the temperature $k_BT$ and $\hbar$. We also investigate how the
cutoff frequency influences this behavior. 

In Fig.~(3) a bifurcation diagram for small values of $\hbar$
and $k_BT$ is shown (small compared with the action and energy
scales of the problem). There is, as before, some coexistence of attractors.
These are included using simulation of the map over many different
initial conditions. 
As expected, large segments of this diagram
are almost equal to the classical counterpart (Fig.~(1)).
There are however, four very important differences: first
of all, and quite visible in the figure, there are chaotic
regions that appear far below the classical threshold
of chaos ($V_0=5.69287$). One of them, at $V_0 \approx 4$,
is associated to the second
bifurcation of the classical diagram. It shows a loss of
stability for the quantum system in a parameter sector close to 
a point where the classical system has a zero Lyapunov exponent.
This behavior is shown in Figs.~(4), which shows the bifurcation
diagram, and (5), which compares the largest Lyapunov
exponent for the quantum and classical maps in this region.
Another appears for $V_0$ around 2.5 and seems to terminate a stable
fixed point branch. A similar one appears for $V_0\approx 5.55$).

Second, and also apparent in Figs.~(4) and (5), is the appearance of a
Hopf bifurcation just before the chaotic threshold for the quantum
map. These transitions are not allowed for the classical system,
since is dissipative and two--dimensional. Here however, they are
preferred over the standard period doubling cascade. We have verified
the appearance of simple non-chaotic cycle attractors, typical of Hopf bifurcations,
for this transition. It should be noticed that Hopf bifurcations
are also present for all other transitions from periodicity to
chaos in this map, of which four are visible in Fig.~(3) (
$V_0 \approx 2.5$, 
$V_0 \approx 4$, 
$V_0 \approx 5.55$ and 
$V_0 \approx 5.7$.) 

The third  main difference between the two bifurcation diagrams
is given by the fact that, although the system is still dissipative,
there seems to be no clear bound for its variables. Here we are showing
$Q$ only for a finite interval, but the numerical simulations generate
values of $Q$ as high as several thousands. This peculiarity affects
some chaotic regions but not all of them.

Finally, there is no evidence of periodic windows or band mergings inside
the chaotic region. There seems to be no topological changes in the
chaotic attractor once the last threshold is crossed. Also,
a great loss of detail can be clearly inferred from the bifurcation
diagram. This is of course even more evident in the attractors
themselves, as seen in the example given in Fig.~(6).
This attractor (at least its projection
on the $Q-P$ plane) still resembles the typical attractor for the
classical map. Here we are also cropping
the figure.
  
Most of these differences are probably associated with the increase
in dimensionality of the map, resulting from
the use of the EVM. As mentioned, a Hopf bifurcation would not
be possible in the classical system.
The behavior of the quantum map, for higher values of $\hbar$ and/or $k_BT$,
is similar to the one already described, except for an
expected growth of the chaotic strips.

It is  also worth mentioning that the attractor (at least its projection
on the $Q-P$ plane) still resembles the typical attractor for the
classical map (see Fig.~(6)). It of course becomes diffused in the plane,
since it is just a projection. Notice that we are strongly cropping
this figure, since values of $P$ and $Q$ spread up to the thousandths.

The changes we have mentioned up to now signal qualitative
differences between classical and quantum maps. Besides these,
we have also studied the evolution of bifurcation points and
chaotic thresholds due to changes in $\hbar$ and $k_BT$, assuming
always $\hbar\ne 0$, $k_BT\ne 0$.
We have obtained  the bifurcation point for the last Hopf transition
and the last chaotic threshold for different values of $k_BT$ and $\hbar$.
In all cases it is found that both points are shifted towards lower values
of $V_0$ with increase of either $k_BT$ or $\hbar$, agreeing with previous
results \cite{graham_tel,cerdeira1}, which  deduce this behavior from
the reduction of quantization and temperature effects  to an effective
noise. This shift seems to have quadratic scaling in both parameters,
in difference with the non--integer exponent change given in
\cite{graham_tel,cerdeira1}. This is not surprising taking into
account the strong truncation in the power series we have performed. 

The temperature effects can be seen in Fig.~(7),
where we plot the value of the potential parameter
$V_0$ for which the systems enters the chaotic regime, as function
of the temperature $k_BT$. 
The different curves correspond
to fixed values of $\hbar$. The solid line set of curves correspond
to the beginning of the Hopf bifurcation where $\lambda = 0$,
and the dashed ones are for
the beginning of fully developed chaos. 
Note that even when this curves are
not as smooth as the first ones, the general tendency is maintained
for all of them.
It can be seen how the value of $V_0$
decreases as $k_BT$ is raised, that is, keeping everything else constant,
the greater the temperature, the earlier chaos appears in the system.
The cutoff frequency has some effect on the curves, but the 
general behavior is the same independently of the value of $\omega_c$.
The influence of the cutoff frequency is more clearly seen
in Fig.~(8), which shows the same two transitions, for fixed $k_BT$ and
variable $\hbar$, and in
Fig.~(9) where it is clear that an increase in the cutoff has a similar
effect as augmenting the $\hbar$ or the temperature, {\em i.\ e.},
it shifts the thresholds for chaos towards lower values of $V_0$.
The scaling of the threshold and bifurcation values with $k_BT$
is quadratic as mentioned before. This can be seen in Fig.~(10).
This is probably due to the truncation to second order the whole EVM.

A very similar behavior is obtained when we change the value of $\hbar$.
In Fig.~(11), the evolution of bifurcation and threshold points
is shown for
several fixed values of $k_BT$. The tendency as before is
to have an early onset of chaos for larger values of $\hbar$. However,
a peculiar phenomenon appears here: for small values of the cutoff parameter,
this tendency is reversed, and the system gets stabilized when
$\hbar$ begins to grow, see Fig.~(12). This anomalous behavior does not
show up when we change $k_BT$.

\section{Conclusions and Final Comments}

The kicked harmonic oscillator with dissipation has been quantized
by means of the Quantum Langevin Equation formalism, truncating
the moment expansion up to second order. We obtained the evolution
maps for the expectation values $Q$ and $P$ and the second order
fluctuations $(\delta Q)^2$, $(\delta P)^2$ and $\,\delta Q\,\delta P
+\delta P\,\delta Q$. The most important result we find is that chaotic
behavior appears in a region of parameter space far below the
classical threshold for chaos. This agrees with the result reported
by Liu and Schieve \cite{schieve}, who also found chaotic behavior
in  a system whose classical version is periodic.
We also find that the change in
dimensionality in the map changes the type of bifurcations in
the dynamics, as shown by the appearance of Hopf bifurcations
on the quantized system. It should be noticed that this is
not an artifact of the EVM, since any representation of a 
quantum system via $c-$numbers (wavefunctions in configuration
space, elements of the density matrix in any given base, etc.)
is infinite dimensional. Therefore,
the reduction of a quantum dissipative system to the corresponding
classical system plus some noise ignores the possible effects of
this change in dimensionality.

Once we are in a finite $k_BT$ and $\hbar$ domain, a continuous
displacement of chaotic thresholds and bifurcation points towards
lower values of $V_0$ is observed, as $k_BT$ and $\hbar$ are increased.
This agrees with most the results obtained up to now in this
field, and goes to confirm the rule that states that {\em quantization
of Hamiltonian systems eliminates chaos, but dissipative systems become
more chaotic when they are quantized}. This is due to the loss
of coherence brought in by the dissipative environment. By now,
there are some experimental confirmations of this rule \cite{exp}.
Here we should comment that irregularities observed in  the curves
for the threshold of chaos in Figs.~(7) and (11) are due to the
uncertainties incurred in the numerical evaluation of the largest
Lyapunov exponent for the system.

We are aware, of course, that a second order expansion is not
sufficient to reproduce the behavior of the
infinite dimensional EVM. Hoverer, we doubt that the inclusion or higher
order corrections will restore periodic behavior to these new chaotic regions,
although
it is of course expected that some extra phenomenology will appear
with the inclusion of these terms.

We have introduced a cutoff frequency $\omega_c$ in order to
perform some of the integrals. These cutoff frequency is expected
to come from the dimensions of the physical system in a real
experiment.  The existence of a cutoff implies a breakdown on the Markovian
condition, but, as long as $\omega_c \gg \Omega$, assuming that
the system is Markovian is not a very strong approximation.
In our simulations, we find that this is not a very important
parameter except when the obviously inconsistent case  when it takes
very low values.

{\bf Acknowledgments}: M.\ R.\ wants to thank
Cinvestav-IPN and the Fondo Yucat\'an for partial support. 
G.\ P.\ acknowledges support from CONACyT, under grant 40726-F.

\appendix{\bf Appendix A: Expectation Values Mapping}

If we are given the collection of moments
\begin{eqnarray}
\langle p^m x^n \rangle &=& {\mbox{\rm Tr}}
( \mbox{\boldmath$\rho$} 
\mbox{\boldmath$p$}^n 
\mbox{\boldmath$x$}^m ) \\
                     &=&\int dx\,\langle x| 
\mbox{\boldmath$\rho$} 
\mbox{\boldmath$p$}^n 
\mbox{\boldmath$x$}^m |x\rangle,
\end{eqnarray}
$\mbox{\boldmath $\rho$}$ 
being the density matrix of the system, we can construct
the function
\begin{eqnarray}
C(\alpha,\beta)&=&\sum_{n,m=0}^\infty\frac{(i\alpha)^m(i\beta)^n
               \langle p^mx^n\rangle}{m!\, n!},\\
               &=&\sum_{n,m=0}^\infty\int dx\,\langle x|
\mbox{\boldmath$\rho$} 
\frac{ (i\alpha)^m(i\beta)^n}{m!\, n!} \,
\mbox{\boldmath$p$}^n 
\mbox{\boldmath$x$}^m, 
|x\rangle,\\
               &=&\int dx\,\langle x|
\mbox{\boldmath$\rho$} \,
\exp{(i\alpha \mbox{\boldmath$p$})}
\exp{(i\beta \mbox{\boldmath$x$})} 
|x\rangle.
\end{eqnarray}
This can also be written as
\begin{eqnarray}
C(\alpha,\beta)&=&\int dx\,dp\,\langle x|
\mbox{\boldmath$\rho$} 
\exp{(i\alpha \mbox{\boldmath$p$})}
|p\rangle\langle p|
\exp{(i\beta \mbox{\boldmath$x$})}
|x\rangle,\\
&=&\frac{1}{\sqrt{2\pi\hbar}}\int dx\,dp\,\langle x|
\mbox{\boldmath$\rho$} 
|p\rangle
e^{i\alpha p+i\beta x-ipx/\hbar}.
\end{eqnarray}
If we Fourier-transform the above equation we get
\begin{eqnarray}
\int d\alpha\,d\beta\,e^{-i\alpha p'}e^{-i\beta x'}C(\alpha,\beta)&=&
\frac{4\pi^2}{\sqrt{2\pi\hbar}}\int dx\,dp\,\langle x|
\mbox{\boldmath$\rho$} 
|p\rangle
e^{-ipx/\hbar}\delta(p-p')\delta(x-x'),\\
&=&\frac{4\pi^2}{\sqrt{2\pi\hbar}}\langle x'|
\mbox{\boldmath$\rho$} 
|p'\rangle
e^{-ip'x'/\hbar}.
\end{eqnarray}
Solving for $\langle x'|
\mbox{\boldmath$\rho$} 
|p'\rangle$ and changing the notation
leads to
\begin{equation}
\langle x|
\mbox{\boldmath$\rho$} 
|p\rangle=\frac{\sqrt{2\pi\hbar}}{4\pi^2}e^{ipx/\hbar}
\int d\alpha\,d\beta\,e^{-i\alpha p}e^{-i\beta x}C(\alpha,\beta).
\end{equation}
If we want to get 
$\mbox{\boldmath$\rho$}$ in a position basis, 
we can take the left hand side of the above equation and do the
following
\begin{equation}
\int dx'\,\langle x|\mbox{\boldmath$\rho$}|x'\rangle\langle x'|p\rangle=\frac{1}
{\sqrt{2\pi\hbar}}\int dx'\,\langle x|\mbox{\boldmath$\rho$}|x'\rangle e^{ipx'/\hbar},
\end{equation}
multiply both sides by $e^{-ipx''/\hbar}$ and integrate over $p$
\begin{equation}
\frac{1}{\sqrt{2\pi\hbar}}\int dx'\,dp\,\langle x|\mbox{\boldmath$\rho$}|x'\rangle
e^{ip(x'-x'')/\hbar}
=\sqrt{2\pi\hbar}\langle x|\mbox{\boldmath$\rho$}|x''\rangle,
\end{equation}
and obtain the desired result
\begin{equation}
\langle x|\mbox{\boldmath$\rho$}|x''\rangle=\frac{1}{4\pi^2}\int dp\,e^{ip(x-x'')/\hbar}
\int d\alpha\,d\beta\,e^{-i\alpha p-i\beta x}C(\alpha,\beta).
\end{equation}
This can also be put in the form
\begin{equation}
\langle x|\mbox{\boldmath$\rho$}|x'\rangle=\frac{1}{2\pi}\int d\beta\,
C\left(\frac{x-x'}{\hbar},\beta\right)e^{-i\beta x}.
\end{equation}

\appendix{\bf Appendix B: Expectation values of the inhomogeneous terms}

In order to calculate the 5 maps for $Q$, $P$ and its three
fluctuations, we will need the following terms
\begin{eqnarray}
\langle\mbox{\boldmath$f$}\cdot\mbox{\boldmath$f$}\rangle&\equiv&f\cdot f,\\
\langle\mbox{\boldmath$h_n$}\cdot\mbox{\boldmath$h_n$}
\rangle&\equiv&h_n\cdot h_n,\\
\langle\mbox{\boldmath$f$}\cdot\mbox{\boldmath$h_n$}+
\mbox{\boldmath$h_n$}\cdot\mbox{\boldmath$f$}\rangle&\equiv&
f\cdot h_n+h_n\cdot f.
\end{eqnarray}
We make the following definitions
\begin{eqnarray}
\mbox{\boldmath$I$}_s(t,\omega_0,\gamma)&=&\frac{1}{\omega_0}\int_{t_0}^t dt'
e^{-\gamma(t-t')}\sin\omega_0(t-t')\mbox{\boldmath$F$}(t'),\\
\mbox{\boldmath$I$}_c(t,\omega_0,\gamma)&=&\frac{1}{\omega_0}\int_{t_0}^t dt'
e^{-\gamma(t-t')}\cos\omega_0(t-t')\mbox{\boldmath$F$}(t').
\end{eqnarray}
The quantities we need are now expressed as
\begin{eqnarray}
f\cdot f&=&\frac{1}{\omega_0^2}I_s\cdot I_s,\\
h_n\cdot h_n&=&\frac{1}{\omega_0^2}\left[(\gamma+V''(Q_n))^2I_s\cdot
I_s+\omega_0^2I_c\cdot I_c-\omega_0(\gamma+V''(Q_n))(I_s\cdot I_c+
I_c\cdot I_s)\right],\\
f\cdot h_n+h_n\cdot f&=&\frac{1}{\omega_0^2}\left[-2(\gamma+V''(Q_n))
I_s\cdot I_s+\omega_0(I_s\cdot I_c+ I_c\cdot I_s)\right],
\end{eqnarray}
and the integrals are explicitly given by
\begin{eqnarray}
\label{I1}
I_s\cdot I_s&=&\frac{2\gamma\hbar}{\pi}\int_0^\infty d\omega\,
\omega\coth\left(\frac{\hbar\omega}{2k_BT}\right)G_{ss}(\omega,\omega_0,
\gamma,\tau),\\
\label{I2}
I_c\cdot I_c&=&\frac{2\gamma\hbar}{\pi}\int_0^\infty d\omega\,
\omega\coth\left(\frac{\hbar\omega}{2k_BT}\right)G_{cc}(\omega,\omega_0,
\gamma,\tau),\\
\label{I3}
I_s\cdot I_c+ I_c\cdot I_s&=&\frac{2\gamma\hbar}{\pi}\int_0^\infty d\omega\,
\omega\coth\left(\frac{\hbar\omega}{2k_BT}\right)G_{sc+cs}(\omega,\omega_0,
\gamma,\tau),
\end{eqnarray}
where 
\begin{eqnarray}
\label{g1}
G_{ss}(\omega,\omega_0,\gamma,\tau)&=&\int_0^\tau dx\,\int_0^\tau dy\,
e^{-\gamma(x+y)}\cos\omega(x-y)\sin\omega_0x\sin\omega_0y,\\
\label{g2}
G_{cc}(\omega,\omega_0,\gamma,\tau)&=&\int_0^\tau dx\,\int_0^\tau dy\,
e^{-\gamma(x+y)}\cos\omega(x-y)\cos\omega_0x\cos\omega_0y,\\
\label{g3}
G_{sc+cs}(\omega,\omega_0,\gamma,\tau)&=&\int_0^\tau dx\,\int_0^\tau dy\,
e^{-\gamma(x+y)}\cos\omega(x-y)\sin\omega_0(x+y).
\end{eqnarray}

The integrals (\ref{g1}-\ref{g3}) are evaluated exactly, and give
expressions that go like $\omega^{-2}$ for large $\omega$.  With them,
and the given values of $\omega_0$, $\gamma$, $\tau$, $k_BT$ and $\hbar$,
integrals (\ref{I1}-\ref{I3}) are done numerically, including a cutoff
factor $e^{-\omega/\omega_c}$ to control the logarithmic
divergence. A Taylor expansion of $\coth$ is
needed for $\omega_0\to 0$.


\centerline{FIGURE CAPTIONS}

\noindent Figure 1. Bifurcation diagram for the classical system for
$1<V_0<9$. Here we have
set $\gamma=0.03, \tau=10$, and use 100 random initial
conditions for each value of $V_0$.

\noindent Figure 2. Attractor for the classical system using the same parameters as
in Fig.~(1), and a value of $V_0=8.0$.

\noindent Figure 3. Bifurcation diagram for the quantized system. We use the
same values of $\gamma$ and $\tau$ as in the classical system
with a cutoff frequency $\omega_c=25$.
Here 100 random initial conditions are used, and $\hbar=k_BT=0.0002$.
Only the central part of the range in $Q$ is shown for clarity.

\noindent Figure 4. Magnification of the quantum bifurcation diagram in the second
chaotic region, showing the presence of the Hopf bifurcation.

\noindent Figure 5. Comparison of the largest Lyapunov exponents as function of
$V_0$. The dotted line corresponds to the classical case,
which undergoes a period doubling bifurcation. In the
quantum case (solid line) this is replaced by a Hopf bifurcation followed by a
chaotic strip.

\noindent Figure 6. Attractor for the quantum system using the same parameters
as in Fig.~(3), and a value of $V_0$ of 8. Only the central part
of the ranges in both $P$ and $Q$ are shown for clarity.

\noindent Figure 7. Plot showing the shift of the chaotic thresholds (dotted line)
and the  Hopf bifurcation (solid line), with respect to
$k_BT$,
for two values of the cutoff frequency. Each curve within any of the 
four bundles
corresponds to a different value of $\hbar$. Starting from the upper curve
the values of $\hbar$ are 0.0005, 0.002, 0.004, 0.006, 0.008 and 0.01. 

\noindent Figure 8. Effect of the cutoff frequency on the chaotic threshold
near $V_0=3.92$.
Values for $k_BT$ and $\hbar$ are fixed at 0.0001 and 0.0005,
respectively.

\noindent Figure 9. Beginning of the Hopf transition as a
function of $k_BT$ for $\hbar=0.006$. with different values for
the cutoff frecuency, $\omega_c=1,5,25$ and 125, with $\omega_c$
augmenting downwards.

\noindent Figure 10. Logarithmic plot of the beginning of the Hopf transition
as function of $k_BT$. The thin line has slope 2.

\noindent Figure 11. Plot showing the shift of the chaotic thresholds (dotted line)
and the  Hopf bifurcation (solid line), with respect to
$\hbar$,
for two values of the cutoff frequency. Each curve within any of the 
four bundles
corresponds to a different value of $k_BT$. Starting from the upper curve
the values of $k_BT$ are 0.0005, 0.002, 0.004, 0.006, 0.008 and 0.01. 

\noindent Figure 12. Anomalous behavior of a chaotic threshold (upper
bundle) and Hopf transition (lower bundle) for a low value
of the cutoff frequency, as function of $\hbar$. Starting from the upper curve
the values of $k_BT$ are 0.0005, 0.002, 0.004, 0.006, 0.008 and 0.01.

\end{document}